# Electrical characterization of all-epitaxial Fe/GaN(0001) Schottky tunnel contacts

Sergio Fernández-Garrido,[a)] Kai U. Ubben, Jens Herfort, Cunxu Gao,[b)] and Oliver Brandt
*Paul-Drude-Institut für Festkörperelektronik, Hausvogteiplatz 5–7, D-10117 Berlin, Germany*

We analyze the properties of Fe Schottky contacts prepared *in situ* on n-type GaN(0001) by molecular beam epitaxy. In particular, we investigate the suitability of these epitaxial Fe layers for electrical spin injection. Current-voltage-temperature measurements demonstrate pure field emission for Fe/GaN:Si Schottky diodes with [Si] = $5 \times 10^{18}$ cm$^{-3}$. The Schottky barrier height of the clean, epitaxial Fe/GaN interface is determined by both current-voltage-temperature and capacitance-voltage techniques to be $(1.47 \pm 0.09)$ eV.

The ferromagnet/semiconductor (FM/SC) Fe/GaN heterostructure is potentially attractive for spintronic applications because of the high Curie temperature of Fe well above 300 K[1] and the long electron spin relaxation time in GaN.[2,3] An essential requirement for spin injection, however, is the formation of a tunnel barrier between the ferromagnetic metal and the semiconductor.[4] Such a tunnel barrier may be realized by a Schottky contact operating in reverse bias.[5,6] An unexpectedly low Schottky barrier height was measured in Ref. 7 for Fe on GaN, but was suggested to be a consequence of the oxidized interface which prevents an intimate contact between the materials.

For spin injection, the constituent materials of the FM/SC hybrid structures are required to be crystalline and pure, and the interface between them should be intimate and abrupt. In previous work by us, we have used molecular beam epitaxy (MBE) to prepare Fe films on GaN(0001) satisfying these conditions.[8,9] The films were not only epitaxial, but the Fe/GaN interface was found to stay structurally and chemically abrupt up to temperatures of 700 °C.[8] The electrical properties of these all-epitaxial Fe/GaN junctions, however, have yet to be explored.

This letter reports on the morphological, magnetic, and electrical properties of all-epitaxial, semitransparent Fe contacts on GaN(0001) suitable for spin-light emitting diodes. Detailed electrical measurements demonstrate that field emission (FE) prevails for Fe/GaN:Si Schottky diodes with [Si] = $5 \times 10^{18}$ cm$^{-3}$. The intrinsic Schottky barrier height (SBH) of Fe/GaN is determined to be $(1.47 \pm 0.09)$ eV.

The Fe/GaN:Si heterostructures were grown in a custom-built MBE system equipped with a radio-frequency N$_2$ plasma source for active N, and solid-source effusion cells for Ga and Fe. Commercially available GaN(0001) templates grown by hydride vapor phase epitaxy (HVPE) on Al$_2$O$_3$ were used as substrates. Si doped and non-intentionally doped (nid) n-GaN layers with a thickness of 0.5 $\mu$m were grown at 690 °C at the boundary between the Ga-droplet and intermediate Ga-rich growth regimes.[10] The Si concentrations of the GaN:Si layers were $5 \times 10^{17}$ and $5 \times 10^{18}$ cm$^{-3}$ as determined by secondary ion mass spectrometry (SIMS), and the residual electron concentration of the nid GaN layer was estimated to be in the low $10^{16}$ cm$^{-3}$ range. After GaN growth, the Ga adlayer was desorbed and an 8.5 nm thick, semitransparent Fe film was epitaxially grown at 50 °C (for further details see Refs. 8 and 9). The surface morphology of this film was examined by atomic force microscopy (AFM). Both the in-plane and out-of-plane Fe magnetization loops were recorded at 300 K in a superconducting quantum interference device magnetometer (SQUID).

Figure 1 displays the in-plane and out-of-plane magnetization loops of the as-grown Fe film at 300 K. The magnetization $M$ is normalized to the saturation magnetization $M_s$ which is close to that of thick Fe films when taking into account the thickness of the surface oxide.[8] The out-of-plane magnetization loop is also virtually identical to that observed for bulk-like Fe. In contrast, the in-plane magnetization exhibits a pronounced anisotropy as visible in the bottom right

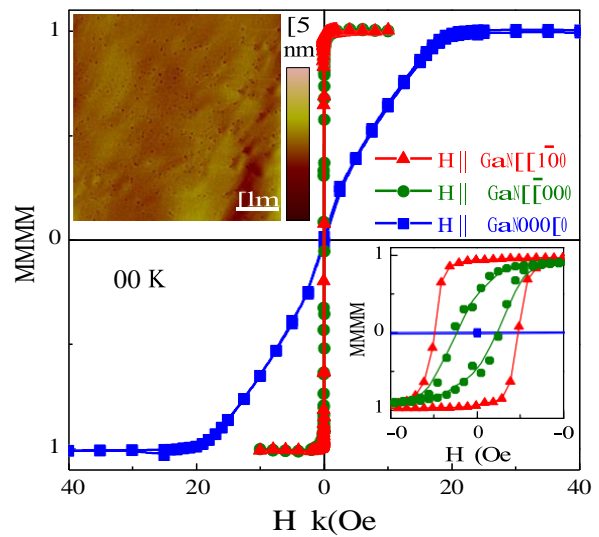

FIG. 1. (Color online) Magnetization loops obtained at 300 K with $H$ parallel to the $[11\bar{2}0]$, $[1\bar{1}00]$, and $[0001]$ directions of GaN. The solid symbols are the experimental data and the lines guides to the eye. The bottom right inset shows the in-plane hysteresis loops for low magnetic fields. The upper left inset shows an AFM micrograph ($5 \times 5$ $\mu$m$^2$) of the as-grown Fe layer.

---

[a)]Author to whom correspondence should be addressed. Electronic mail: garrido@pdi-berlin.de
[b)]Present address: School of Physical Science and Technology, Lanzhou University, Lanzhou 730000, P. R. China

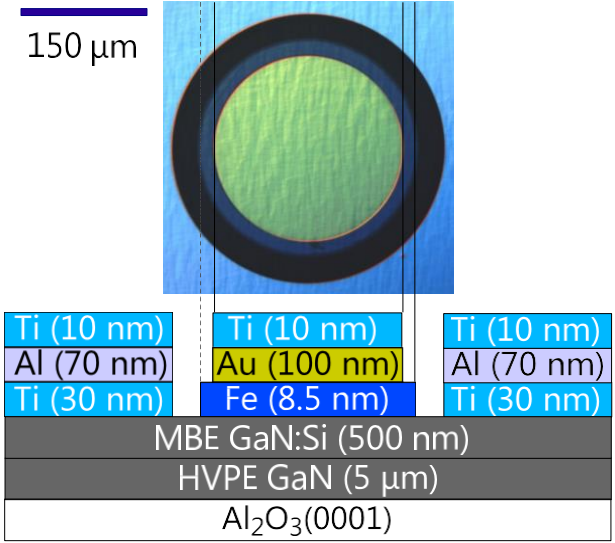

FIG. 2. (Color online) Optical micrograph (top view) and schematic structure of an Fe/GaN:Si Schottky diode.

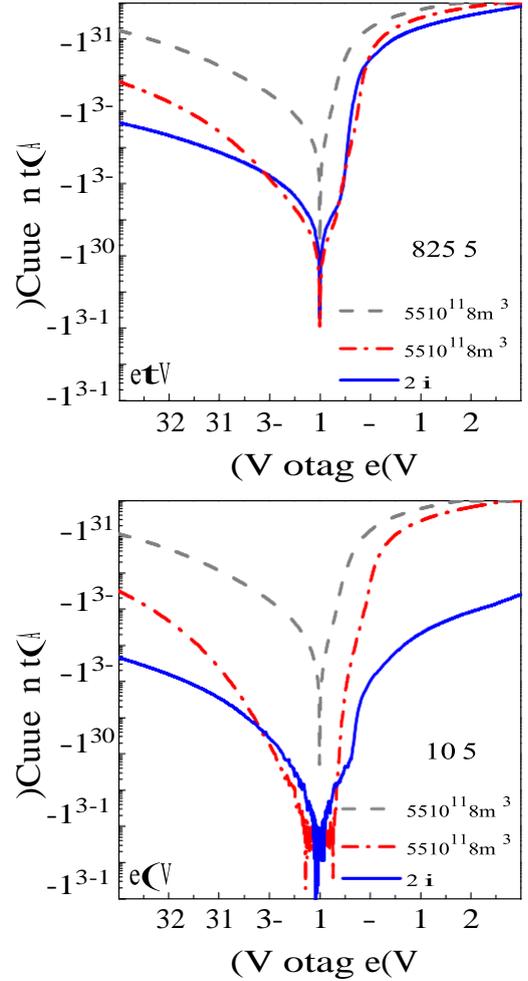

FIG. 3. (Color online) Current-voltage characteristic of three Fe/GaN Schottky diodes with different doping levels at 295 K (a) and 80 K (b).

inset of Fig. 1. This anisotropy is not observed for thick films[8] and is thus presumably an interface-induced effect.

The upper left inset in Fig. 1 shows a $5\times5$ $\mu m^2$ AFM micrograph of the as-grown Fe/GaN heterostructure. The Fe film exhibits a root-mean-square roughness below 1 nm. The dark spots with an areal density of $\approx 1.3 \times 10^9$ cm$^{-2}$ stem from threading dislocations originating in the GaN template and intersecting the MBE-grown GaN layer.

Figure 2 shows an optical micrograph and a schematic of the Fe/GaN Schottky diodes. The planar Schottky diodes, consisting of a solid Fe disk (∅ 250 $\mu$m) and an extended ohmic contact, were fabricated using standard optical lithography, wet etching, electron-beam evaporation, and lift-off techniques. To improve the linearity of the ohmic contact, the diodes were annealed at 480 °C for 10 min in an N$_2$ atmosphere. The electrical properties of these diodes were investigated by current-voltage-temperature (I-V-T) and room temperature capacitance-voltage (C-V) techniques. Both types of measurements were carried out in the dark.

Figures 3(a) and (b) display the I-V characteristics of the diodes under investigation at 295 and 80 K, respectively. The nid diode shows a clear rectifying behavior at 295 K with an exponential dependence on the applied voltage for a forward bias between 0.4 and 0.7 V. At higher voltages, the dependence changes to a linear one due to a series resistance of d$V$/d$I \approx 28$ Ω. Under reverse bias, the current is more than three order of magnitude lower than under forward bias exhibiting a value as low as 0.05 mA at $-4$ V. For the Si doped diodes, the current under both forward and reverse bias is higher and increases with Si concentration. The comparison of the I-V curves at 295 and 80 K reveals that the Si doped diodes are not as sensitive to the temperature as the nid diode. In fact, for the diode with the highest Si concentration the current is essentially independent of the temperature. The differences between the I-V characteristics of the nid and the Si doped diodes can be understood by taking into account that thermionic field emission (TFE) and FE transport mechanisms are enhanced in the Si doped diodes because the depletion region becomes narrower.[11] Considering that for electrical spin injection a high reverse current is desired and pure FE transport is required to circumvent the conductivity mismatch in metal/semiconductor junctions,[4] we focus hereafter on the analysis of diode with the highest Si concentration ($5\times10^{18}$ cm$^{-3}$).

To elucidate the transport mechanism in this diode, we first analyze the forward I-V-T characteristic between 20 and 295 K [Fig. 4(a)] within the voltage range for which an exponential behavior is observed [see Fig. 4(b)]. Under forward bias, the relation between the current and the voltage for a Schottky diode can in general be written as:

$$I_F = I_S \exp(eV/nkT), \quad (1)$$

where $I_s$, $e$, $n$, and $k$ are the saturation current, the electron charge, the ideality factor, and the Boltzmann constant, respectively. If FE is dominant, as required for spin injec-tion, $n = E_{00}/kT$, and $I_s$ is a function of the SBH $\varphi_B$ and the characteristic tunneling energy $E_{00}$ whose value can be calculated to be 31 meV for the diode under investigation.[12]

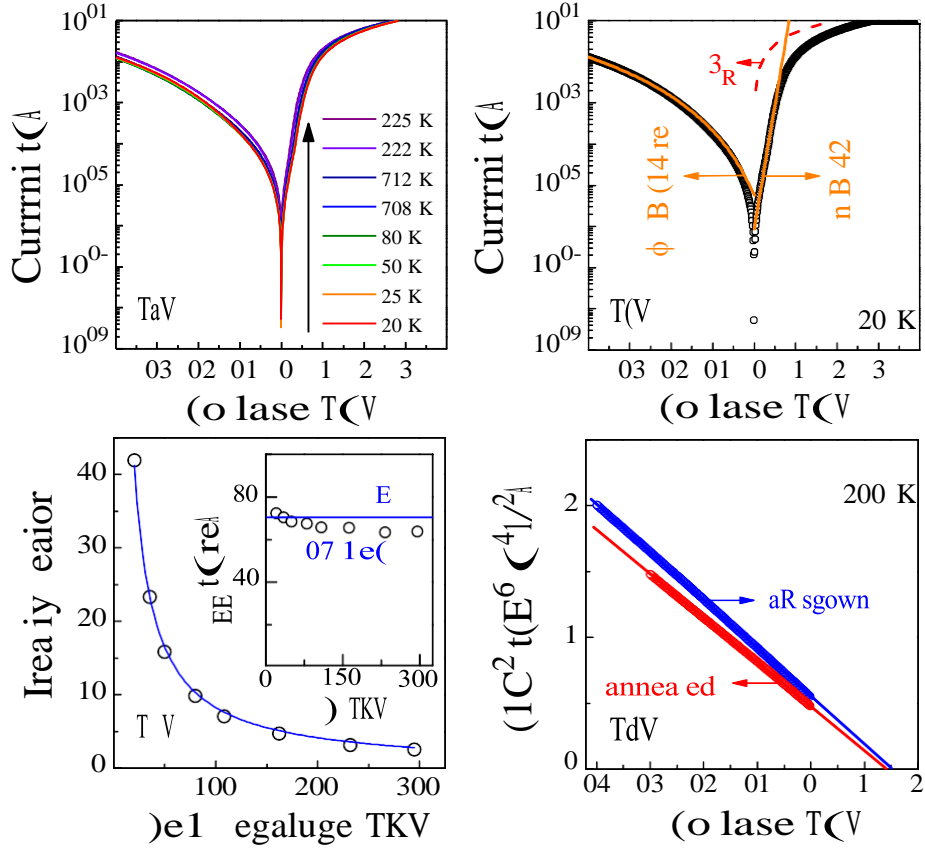

FIG. 4. (Color online) (a) Temperature dependence of the I-V characteristic of the Fe/GaN:Si Schottky diode between 20 and 295 K. (b) Current-voltage characteristics of the Schottky diode at 20 K. The open circles represent the experimental data, the solid lines are the fits to Eqs. (1) and (2), and the dashed line illustrates the effect of the series resistance. (c) Temperature dependence of the forward ideality factor. The solid line shows the fit to $n = E_{00}/kT$. In the inset, the open circles represents the temperature dependence of $E_{00}$, calculated as $nkT$, and the solid line indicates the value of $E_{00}$ derived from the fit. (d) Plot of $1/C^2$ versus $V$ for annealed and as-grown Fe/GaN:Si Schottky diodes with a Si concentration of $5 \times 10^{17}$ cm$^{-3}$.

Figure 4(c) shows the temperature dependence of $n$ derived from the fit of the I-V curves to Eq. (1). The ideality factor decreases with temperature, and the experimental data can be properly fit to $n = E_{00}/kT$ with $E_{00} = 71$ meV. Hence, $E_{00}$ is indeed essentially constant as shown in the inset in Fig. 4(c).

The temperature independence of $E_{00}$ demonstrates that FE dominates for the given Si doping level.[12] The fact that the experimental value of $E_{00}$ is higher than the theoretical one reflects a higher tunneling probability than expected.[12] Such a deviation has also been observed by other groups for GaN Schottky diodes, and is commonly explained in terms of defect-assisted tunneling given that the high density of defects and dislocations present in heteroepitaxial GaN layers may provide additional tunneling paths through the energy barrier.[11,13–16]

Since FE is the dominant transport mechanism under forward bias, it should also govern the transport under reverse bias, the preferred configuration for electrical spin-injection.[4] The I-V curve for $-eV > \varphi_B$ is given by:

$$I_R = AA^* \frac{E_{00}^2(\varphi_B - eV)}{k^2 \varphi_B} \exp\left[-\frac{2\varphi_B^{3/2}}{3E_{00}(\varphi_B - eV)^{1/2}}\right], \quad (2)$$

where $A$ and $A^*$ are the Schottky contact area and the effective Richardson constant of Fe, respectively.[11,12] Figure 4(b) shows, as an example, the fit of the reverse I-V curve obtained at 20 K to Eq. (2) using the value for $E_{00}$ obtained above and treating $\varphi_B$ and $AA^*$ as fit parameters. A satisfactory fit is obtained for $\varphi_B = 1.4$ eV and $AA^* = 4 \times 10^{-6}$ AK$^{-2}$. With increasing temperature, $\varphi_B$ slightly decreases from 1.40 to 1.29 eV, and $AA^*$ scatters within a range of 4–8×10$^{-6}$ AK$^{-2}$. Taking into account that the Schottky contact area is $\approx 5 \times 10^{-4}$ cm$^{-2}$, the effective Richardson constant derived from the fit is 0.8–1.6×10$^{-2}$ Acm$^{-2}$K$^{-2}$.

In order to corroborate the value of the SBH derived above, C-V measurements were performed for two different diodes with a Si concentration of [Si] $= 5 \times 10^{17}$ cm$^{-3}$ (the high tunneling current under reverse bias for the diode with [Si] = $5 \times 10^{18}$ cm$^{-3}$ prevents its characterization by C-V). The first one was prepared as described above, and for the second one, the annealing step was omitted. Figure 4(d) presents the plots of $1/C^2$ versus $V$ at 1 MHz using an excitation voltage of 0.02 V. In both cases, the electron concentration required to fit the data is close to the actual Si concentration determined by SIMS. For the SBH, the fit returns $(1.471 \pm 0.002)$ and $(1.552 \pm 0.003)$ eV for the annealed and the as-grown diode, respectively, and thus an only marginally higher value than the average one derived from the I-V-T measurements above $(1.38 \pm 0.07)$ eV. The difference can be ascribed to image-charge effects.[13] The C-V results therefore compare well with those derived from the analysis of the I-V curves and furthermore evidence that the annealing step does not strongly influence the SBH.

The classical Schottky-Mott theory predicts a SBH for Fe/GaN of 0.6 eV and evidently cannot account for the SBH of these all-MBE Fe/GaN contacts. The SBH measured is also larger than the value of about 1 eV predicted by the theory of metal-induced gap states.[7] In this context, it is interesting to note that we have measured a SBH of 0.6 eV for polycrystalline Fe contacts deposited *ex situ* onto GaN surfaces exposed to air.[17] A similarly small value has been reported by Kampen and Mönch[7] for their non-intimate Fe/GaN contacts. The significantly higher SBH of the epitaxial Fe/GaN contacts measured here may thus be of interest for evaluating theoretical concepts of Schottky contacts. In the present context, we conclude that our results suggest that all-MBE Fe/GaN contacts qualify for spin injection. At sufficiently high doping level, transport occurs by pure FE as desired to circumvent the conductivity mismatch in metal/semiconductor junctions.

The authors would like to thank K.-J. Friedland for fruitful discussions, H.-P. Schönherr for the maintenance of the MBE system, W. Anders, B. Drescher, and W. Seidel for the fabrication of the devices, and M. Ramsteiner for a critical reading of the manuscript.